\documentclass[aps]{revtex4-1}

\usepackage{graphicx}
	\graphicspath{{./Analysis/}}
	\DeclareGraphicsExtensions{.pdf,.png,.jpg}

\newcommand{\degr}{\ensuremath{^{\circ}}}

\newcommand{\eqn}[1]{Eqn.~(\ref{#1})}
\newcommand{\Bo}{\ensuremath{Bo}}
\newcommand{\ABo}{\ensuremath{A_0 Bo}}
\newcommand{\We}{\ensuremath{W\!e}}
\newcommand{\Wecrit}{\ensuremath{W\!e_{\mathrm{crit}}}}
\newcommand{\Cwcrit}{\ensuremath{C_{w,\mathrm{crit}}}}
\newcommand{\Cgcrit}{\ensuremath{C_{\gamma,\mathrm{crit}}}}

\begin{document}

\title{Wind- and Gravity-Forced Drop Depinning}
\author{Edward B. White} \email{ebw@tamu.edu}
\author{Jason A. Schmucker} 
\affiliation{Department of Aerospace Engineering,
Texas A\&M University, College Station, TX 77843, USA}
\date{\today}

\begin{abstract}

Liquid drops adhere to solid surfaces due to surface tension but can
depin and run back along the surface due to wind or gravity forcing.
This work develops a simple mechanistic model for depinning by combined
gravity and high-Reynolds-number wind forcing and tests that model using
water drops on a roughened aluminum surface. On non-inclined surfaces,
drops depin at a constant critical Weber number,
$W\!e_{\mathrm{crit}}=7.9$, for the present wettability conditions. On
inclined surfaces, $W\!e_{\mathrm{crit}}$ decreases linearly with the
product of the Bond number and the width-to-height aspect ratio of the
unforced drop. The linear slope is different in distinct wind- and
gravity-dominated forcing regimes above and below
$W\!e_{\mathrm{crit}}=4$. Contact line shapes and drop profile shapes
are measured at depinning conditions but do not adequately explain the
differences between the two forcing regimes.

\end{abstract}

\maketitle

\section{Introduction}

When a liquid drop rests on a solid surface in the presence of wind
or if the surface is inclined, the drop may remain fixed in place
or depin and run back along the surface. When forcing is low,
surface tension balances the wind and gravity forces and the drop
remains pinned in place. As forcing increases, it eventually exceeds the
maximum pinning force surface can provide and the drop depins and runs
downstream.

Whether a drop depins involves a complex balance of forces
at the three-phase contact line. Contact angle hysteresis is critical
as it provides the drop with a range of metastable configurations and determines
the criterion for depinning of the contact line~\cite{Macdougall-PRSLA-42}.
The advancing contact angle, $\theta_a$, is the maximum angle on
the advancing side of the drop; the receding contact angle, $\theta_r$,
is the minimum contact angle on the receding side. Because the
advancing angle exceeds the receding angle, surface tension exerts
a net force in the direction opposing motion. The maximum pinning
force is largely determined by the difference between the cosines of these
angles, $\Delta(\cos\theta)_{a,r}=\cos\theta_r-\cos\theta_a$.

When a drop rests on a surface inclined at angle $\alpha$, the
downhill force is $\rho_d gV\sin\alpha$, where $\rho_d$ is the drop
density.  Wind forcing involves the interface pressure and viscous
stress imposed on the drop by air. These depend strongly on drop
shape which is affected by contact angle and the impinging flow
field. When a drop is very small or the air (or other fluid) velocity
is low, the forcing on the drop is mainly due to shear stress. When
the drop is large or the air velocity is high, the dynamic pressure
stress imposed by the air becomes dominant. Whether shear stress
or pressure is more important is characterized by a Reynolds number,
$Re=\rho_a U h \big/ \mu_a $, based on the drop height, $h$, a
characteristic air velocity, $U$, at that height, and the viscosity,
$\mu_a$, and density, $\rho_a$ of the air. The present interest is
high Reynolds numbers. In the high-$Re$ regime, flow over a drop
separates, vortices are shed~\cite{Acarlar-JFM-87}, and pressure
fluctuations may cause drop-shape unsteadiness~\cite{White-JFE-08}.

The objective of this work is to develop and test a simple mechanistic
model of drop depinning limits with combined gravity and high-$Re$-wind
forcing. Although both forcing types may act simultaneously on drops, mixed
forcing results do not appear to be addressed elsewhere. Examining mixed
forcing enables a systematic study of whether depinning characteristics
depend on forcing modality. This may provide new insight into how
laboratory depinning studies using tilted plates may or may not apply to
industrial applications in which high-$Re$ forcing is most important.

The paper is organized as follows. After a brief literature review in
Sec.~\ref{background}, a model for depinning limits is developed in
Sec.~\ref{model}. Section~\ref{setup} provides an overview of the
experimental approach using a small tiltable wind tunnel. Drop depinning
results and a discussion of implications for the model are given in
Sec.~\ref{depinning}. Section~\ref{shape} provides data on drop shapes
under different forcing conditions as a potential explanation for the
results presented in Sec.~\ref{depinning}. Conclusions are discussed in
Sec.~\ref{conclusions}.

\section{Background}
\label{background}

Early gravity-forced experiments on inclined surfaces without wind
forcing were by Macdougall and Ockrent~\cite{Macdougall-PRSLA-42},
Bikerman~\cite{Bikerman-JCS-50}, and Furmidge~\cite{Furmidge-JCS-62}.
These experiments revealed that larger contact angle hysteresis
increases a drop's ability to resist depinning. To capture this
behavior, Macdougall and Ockrent proposed a depinning model equation
equivalent to \begin{equation} \alpha_{\mathrm{crit}} = \sin^{-1}
\left(\frac{\gamma\,w\,\Delta(\cos\theta)_{a,r}}{\rho_d \,g \,V}
\right) \label{Macdougall} \end{equation} in which $\gamma$ is the
surface tension, $V$ and $w$ are the drop volume and width, and
$\alpha_{\mathrm{crit}}$ is the angle at which the pinning force
equals the gravity force in the downhill direction.  Bikerman
observed modifications to the contact line and contact angle as the
critical inclination is approached. These changes are required to
maintain equilibrium between the gravity and surface tension before
the gravity force exceeds the maximum available pinning force.

Since that early work, multiple researchers have studied the problem
of drops depinning on inclined planes. Seminal analytical studies were
by Dussan~V. and Chow~\cite{Dussan-JFM-83,Dussan-JFM-85}. The former
study considered the depinning problem for small contact angles in the
lubrication limit. The latter relaxed the lubrication assumption and
only required $\Delta(\cos\theta)_{a,r}$ to be small. More recent
studies are by Qu\'er\'e et al.~\cite{Quere-Langmuir-98}, ElSherbini and
Jacobi~\cite{ElSherbini-JCIS-06}, Berejnov and Thorne~\cite{Berejnov-PRE-07},
Chou et al.~\cite{Chou-Langmuir-12}, and others.
Qu\'er\'e et al.\ develop and test a depinning model for drops
with small contact-angle hysteresis in the low-Bond-number limit
over a wide range of advancing and receding contact angles.
ElSherbini and Jacobi provide an analytical model that predicts a
critical angle based on the receding contact angle. As tilt angle
increases, multiple critical angles exist as drops transition
between different metastable configurations prior to
depinning~\cite{Berejnov-PRE-07,Chou-Langmuir-12}. Berejnov and
Thorne show experimental results on modifications to the contact
line shape and the uphill and downhill contact angles as inclination
increases. Chou et al.\ observe the same behavior in experiments
and numerical simulations.

Wind-forced depinning is substantially more complex than the
gravity-forced depinning because the wind profile can be an important
factor. At high Reynolds numbers, wind forcing can also be unsteady
and lead to drop interface oscillations~\cite{White-JFE-08}. Milne
and Amirfazli~\cite{Milne-Langmuir-09} give a review of the wind-forced
depinning literature. Recently, Razzaghi et al.~\cite{Razzaghi-PhysFluids-18}
considered how critical wind velocity is affected when drops are
positioned in closely spaced arrays.

Milne and Amirfazli~\cite{Milne-Langmuir-09} propose a high-$Re$
depinning model that predicts the wind velocity at which aerodynamic
drag, $F_{\mathrm{drag}} = \frac{1}{2} \rho_a U^2\! A\,C_D$, becomes
equal to the maximum drop adhesion force which they characterize
as $F_{\mathrm{adh}} = \gamma k L_b \Delta(\cos\theta)_{a,r}$. The
drag depends on the drop projected area $A$ before wind forcing and
drag coefficient $C_D$ which depends on the drop's volume and
shape. The shape depends on the forcing magnitude, surface tension,
and contact angle limits. The pinning force depends on $L_b$, the
drop base length (diameter) before forcing is applied, the advancing
and receding contact angles, plus a parameter $k$ that accounts for
contact-angle variations about the drop's contact line. Similar
to $C_D$, $k$ depends on parameters that may change as depinning
is approached. Milne and Amirfazli equate the drag and pinning
forces and solve for a critical depinning velocity, $U_{\mathrm{crit}}$
that is proportional to $(L_b/A)^{1/2}$ and involves the advancing
and receding contact angles, surface tension, air density, and the
unknown ratio $k/C_D$.

Working with four different liquid/solid combinations, Milne and
Amirfazli found $U_{\mathrm{crit}}$ depinning data did not readily
fit the $(L_b/A)^{1/2}$ form and instead found a function
$U_{\mathrm{crit}} = a\,\exp\left[b\,(L_b/A)^{1/2}\right]$ to be
more successful. The fitting parameters $a$ and $b$ are different
for each different liquid/solid pair. The fact that $(L_b/A)$ does
not appear to the 1/2 power but instead inside an exponential
indicates that the $k/C_D$ term is sensitive to $(L_b/A)$ plus,
potentially, other parameters. Later analysis by Roisman et
al.~\cite{Roisman-PRE-15} showed Milne and Amirfazli's depinning
data to collapse along $U_{\mathrm{crit}} \sim V^{-1/3}$ curves
which accounts for the variability in the drop size relative to the
shear-layer thickness.

\section{Mixed-Forcing Depinning Model}
\label{model}

The proposed model for depinning limits derives from a simple force
balance similar to the gravity-only model proposed by Macdougall
and Ockrent~\cite{Macdougall-PRSLA-42} plus the wind-only model
suggested by Milne and Amirfazli~\cite{Milne-Langmuir-09}. Detailed
drop shapes, wettablity characteristics, and high-$Re$ unsteadiness
are not explicitly included. Instead, their various effects are
lumped into model coefficients that are experimentally measured.

Using this approach, the pinning force provided by surface tension is
modeled as $F_{\gamma} = C_{\gamma}^*\,\gamma\,w_0$ where
$w_0$ is the initial width of the circular drop contact line and
$C_{\gamma}^*$ is an $\mathcal{O}(1)$ coefficient that accounts for
the contact line shape and the contact angle
distribution about the contact line. This model is similar to the
exact equations developed by ElSherbini and
Jacobi~\cite{ElSherbini-JCIS-06} but lumps the contact line shape
and contact angle variability into the unknown
parameter $C_{\gamma}^*$. The use of $w_0$ is consistent with those authors'
finding that the equivalent drop radius is most appropriate for
that purpose.

Because $C_{\gamma}^*$ depends on drop shape, it increases as forcing
increases to maintain equilibrium. How it does so may depend on the
forcing mode. Depinning occurs when forcing exceeds the maximum
available pinning force. The maximum is expected to be proportional
to $\Delta(\cos\theta)_{a,r}$.

The gravity force on a drop is $F_{\mathrm{grav}} = C_g^*\,\rho_d\,
g\,w_0^2 \, h_0 \sin\alpha$. The initial drop height is $h_0$ and
$C_g^*$ is an $\mathcal{O}(1)$ constant that relates the actual
drop volume to $w_0^2\,h_0$. Because the drop volume is fixed,
$C_g^*$ does not depend on forcing but only the initial drop shape.
At low Bond numbers when drops have the shape of spherical caps,
$C_g^*$ can be computed exactly using the measured drop height and
width. The aerodynamic force is $F_{\mathrm{wind}} = C_w^*\,\rho_a\,
U^2\,w_0\,h_0$ where $U$ is a characteristic wind speed and $C_w^*$
is an $\mathcal{O}(1)$ drag coefficient that depends on the
instantaneous drop shape. $C_w^*$ is expected to change markedly
as air velocity is increased and the drop shape evolves in response
to this forcing.

Combining the force terms into a single equation yields \begin{equation}
C_{\gamma}^*\,\gamma\,w_0 = C_w^*\,\rho_a\,U^2\,w_0\,h_0 + C_g^*\,
\rho_d\,g\,w_0^2\,h_0\,\sin\alpha, \label{dimensional-model}
\end{equation} for a pinned drop up to the maximum value of $C_{\gamma}^*$.
Once the right-hand side exceeds the maximum value,
the drop depins. Equation~\ref{dimensional-model} can be
recast as \begin{equation} C_{\gamma} = C_w\,\We + \ABo\,\sin\alpha
\label{full-model} \end{equation} in which \Bo\ and $\We$ are the
Bond and Weber numbers, $A_0 = w_0\big/h_0$ is the initial drop aspect
ratio, and the unknown coefficients are combined as $C_{\gamma}=C_{\gamma}^*\big/C_g^*$
and $C_w=C_w^*\big/C_g^*$. The $C_g^*$ coefficient is absorbed into
$C_{\gamma}$ and $C_w$ because it does not change as forcing increases.
The Bond number is defined $\Bo=\rho_d\,gh_0^2\big/\gamma$. The
Weber number is defined $\We=\rho_a U^2h_0\big/\gamma$.

Equation~\ref{full-model} reveals the role of the Bond and Weber
numbers in depinning and presents an immediate implication for
gravity-only forcing. The critical depinning angle is \begin{equation}
\alpha_{\mathrm{crit}} = \sin^{-1}\!\left( \frac{\Cgcrit}{\ABo}
\right), \label{critical-tilt}\end{equation} which is essentially
the same as \eqn{Macdougall}\ and a result by ElSherbini and
Jacobi~\cite{ElSherbini-JCIS-06} who find a critical tilt angle
based on the Bond number. The aspect ratio and Bond number are
known based on initial conditions. \Cgcrit\ is unknown and
may depend on the initial and final drop shape which, for gravity-only
forcing, could only depend on drop volume and wettability parameters.
The degree to which $C_a$ depends on volume in gravity-only forcing
can be assessed by conducting gravity-only depinning tests and
evaluating the quality of a fit to the \eqn{critical-tilt}\ model
assuming \Cgcrit\ is constant.

Considering situations with subcritical tilt, \eqn{full-model}\ can
be rearranged to yield the critical Weber number \begin{equation}
\Wecrit = \frac{\Cgcrit}{\Cwcrit} - \frac{1}{\Cwcrit} \ABo\,
\sin\alpha. \label{critical-wind}\end{equation} It is immediately
apparent that the critical Weber number is $\Cgcrit\big/\Cwcrit$
for non-tilted surfaces.  This is a sensible result because a stronger
pinning force requires more wind velocity to depin a drop while a higher
drag coefficient requires less. The value of \Cwcrit\ and the ratio
$\Cgcrit\big/\Cwcrit$ may only be weak functions of Bond number.
How they vary can be assessed using wind-only depinning tests
at various Bond numbers and observing the manner in which the
critical Weber number depends on \ABo. Milne and
Amirfazli~\cite{Milne-Langmuir-09} cite evidence that, at least
over small volume ranges, critical Weber numbers may be constant
for $\alpha=0\degr$. More generally, \eqn{critical-wind}\ suggests
that the critical Weber number is a linearly decreasing function
of \ABo\ if the unknown coefficients are not strong functions of
\ABo.

The sections below present experimental tests of the depinning model as
represented by Eqns.~(\ref{critical-tilt}) and~(\ref{critical-wind}).
The key questions this work aims to address are: Is this simple
mechanistic model a useful representation of wind-forced depinning with
various degrees of surface inclination? If yes, what are the values of
the unknown parameters \Cgcrit\ and \Cwcrit\ and are they functions of
Bond number or other factors? And, finally, can the values of
\Cgcrit\ and \Cwcrit\ be rationalized with respect to contact-line shape and
contact-angle variability in different forcing and Bond-number regimes?

\section{Experimental Setup and Procedures}
\label{setup}

\subsection{Wind Tunnel and Flow Conditions}

The experiments seek to quantify wind- and gravity-forced depinning
limits in the air/water/aluminum system using a small tiltable wind
tunnel. The tunnel was developed by Schmucker~\cite{Schmucker-PhD-12}
and was used previously by Hooshanginejad and
Lee~\cite{Hooshanginejad-PRF-17}. Drops are placed on the floor of
the wind-tunnel test section that consists of a roughened aluminum
substrate. Side- and top-view cameras are included on the tilting
platform to provide the nonintrusive drop-shape measurements developed
by Schmucker and coworkers~\cite{Schmucker-PhD-12,Schmucker-ExpFluids-12}.

\begin{figure}
\includegraphics[width=3.5in]{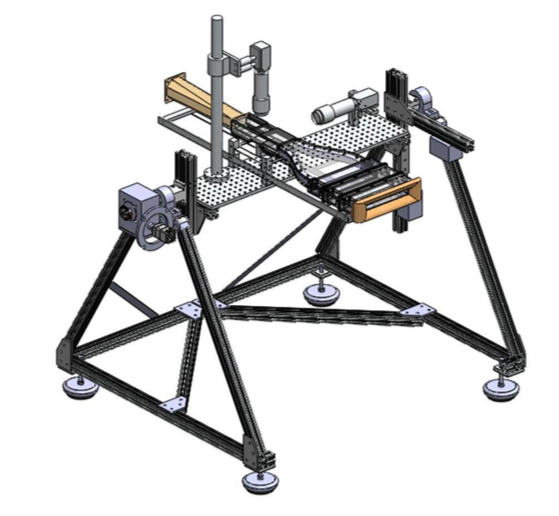}
\caption{Wind tunnel and experimental rig for drop stability experiments}
\label{tunnel}
\end{figure}

The wind tunnel is designed according to typical wind tunnel
paradigms~\cite{Barlow-99}. A schematic is shown in Fig.~\ref{tunnel}.
It is an open-return design with an inlet cross section 25~mm tall
by 200~mm wide. After passing through a honeycomb and two screens
for flow conditioning, a 250-mm-long contraction decreases the cross
section to 25~mm tall by 50~mm wide. Interchangeable surface samples
25~mm wide by 50~mm long fit flush into the tunnel floor. The
sandblasted aluminum surface has a rms surface roughness of
3.26~$\mu$m. Downstream of the test section, the flow passes through
a diffuser and an 80-mm-diameter fan. The wind-tunnel rotation axis
passes through the test-section so the linear acceleration of a
drop is essentially zero while the tunnel is rotated to different
inclination angles. The pressure drop across the contraction is
measured to control the test-section flow velocity. Surface inclination
is measured using a rotary encoder.

\begin{figure}
\includegraphics[width=5in]{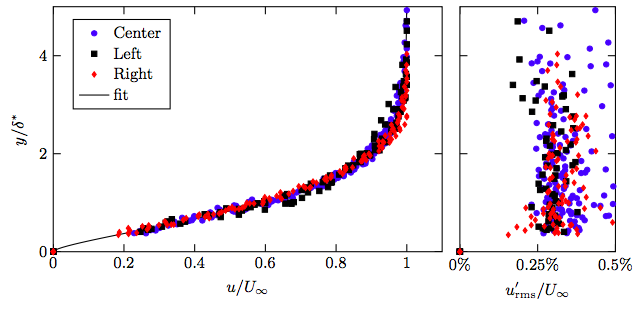}
\caption{Boundary velocity and turbulence intensity profiles}
\label{bl-profiles}
\end{figure}

A hotwire anemometer was used to measure the wind-tunnel floor
boundary layer.  Normalizing the height above the surface, $y$, by
the boundary-layer displacement thickness, $\delta^*$, and the flow
velocity by the freestream velocity, $U_{\infty}$, the data collapse
to a self-similar curve shown in Fig.~\ref{bl-profiles}. The
displacement thickness was found to vary as $\delta^* = a\,
U_{\infty}^{-0.5}$ where $a = 1.7\ \mathrm{mm(m/s)}^{1/2}$. As seen
in Fig.~\ref{bl-profiles}, $u'_{\mathrm{rms}}$ velocity fluctuations
are between 0.2 and 0.5\% of the freestream speed.

\subsection{Test Procedures}

To begin a depinning experiment, the aluminum substrate is cleaned with
acetone. Once it evaporates, an image of the dry aluminum is captured
using the top-view camera. Next, a distilled water drop of a
particular volume, $V$, is applied on the surface. Drops are applied
by hand using a graduated syringe. Careful drop application is
essential to produce nearly circular contact lines with contact
angles close to $\theta_a$. A second top-view image is captured
once the drop is applied. Next, the wind tunnel is brought to a
sub-critical inclination angle at 1\degr\ per second. Once the
target inclination is reached, wind speed is slowly increased until
the critical speed is reached and the drop depins. Top- and side-view
images are captured at multiple subcritical flow speeds.

Depinning is identified using side- and top-view images collected
during the experiments. Depinning is judged to occur at the velocity
or inclination at which motion is first observed on the receding
portion of the contact line. This corresponds to the second of three
depinning events identified by Berejnov and Thorne~\cite{Berejnov-PRE-07}.
Consistent with observations by Berejnov and Thorne, the advancing
side of a drop is usually observed to move first in response to
subcritical forcing. This allows the drop to temporarily achieve a
meta-stable configuration before stronger forcing eventually causes
depinning.

\subsection{Drop Geometry}
\label{drop-geometry}

\begin{figure}
\includegraphics[trim=48 48 160 496]{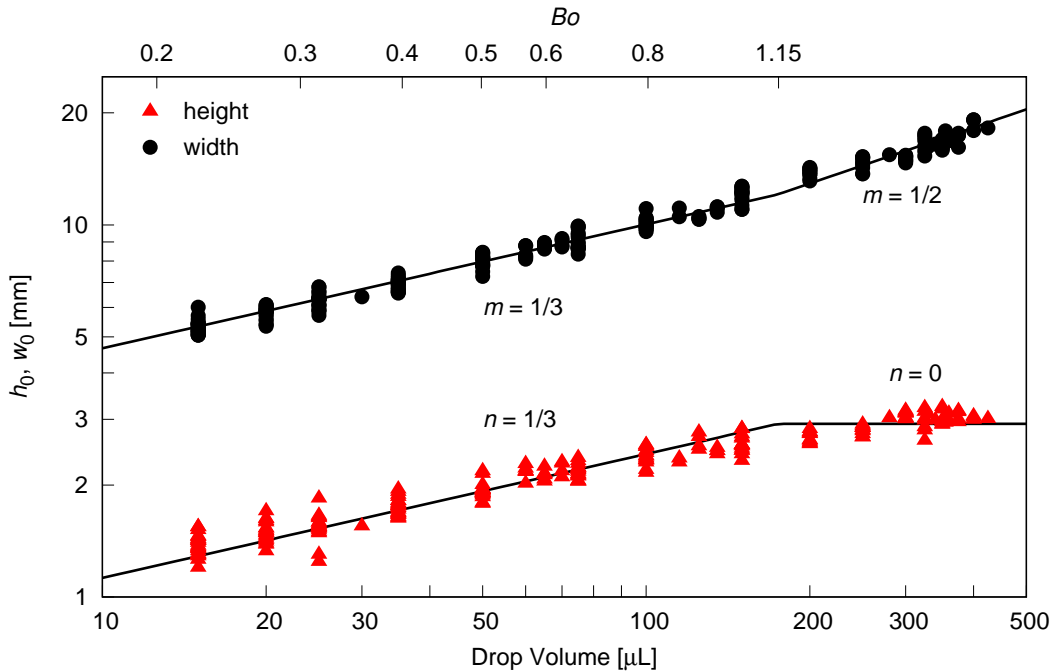}
\caption{Initial drop-dimension measurements}
\label{drop-dimensions}
\end{figure}

Side-view images obtained at $\alpha=0\degr$ and $U_{\infty}=0$~m/s
at the start of each test are used to measure the initial height,
$h_0$, and width, $w_0$, of each drop. Before forcing is applied,
the drop width is the diameter of the nearly circular contact line.
As shown in Fig.~\ref{drop-dimensions}, these measurements both
scale as $V^{1/3}$ at low Bond numbers where drops take the shape
of spherical caps. At large drop volumes, the initial heights approach
a constant value of 2.92~mm, approximately equal to the capillary
length $\ell_c \sim \big( \gamma \big/ \rho_d g
\big)^{1/2} \approx 2.72$~mm. At larger volumes, $w_0$ increases
as $V^{1/2}$.

The data in Fig.~\ref{drop-dimensions} are fit to power-law forms
$h_0\propto V^n$ and $w_0 \propto V^m$ in two ranges with a break
at $V=175\ \mu$L, the cutoff between low and high Bond numbers.
This volume corresponds to $\Bo=1.15$. Although there is actually
a smooth transition between these regimes, a piecewise-continuous
fit successfully models the data within measurement uncertainty.
The break at $175\ \mu$L plus the proportionality constants for
$h_0$ and $w_0$ in the low-Bond regime are the three fit parameters.
Bond numbers corresponding to each volume are shown at the top of
the figure. Because $h_0$ is constant above $\Bo=1.15$, no further
increase in the Bond number occurs at higher volumes. The initial
drop aspect ratio is constant in the low-Bond regime, $A_0=w_0\big/h_0=4.14$.
This result implies that $C_g^* \approx 0.42$ for $\Bo \le 1.15$.
At higher Bond numbers the drop is no longer a spherical cap so $A_0$
increases as $V^{1/2}$ and $C_g^*$ increases.

As the tests proceed, side-view images are used to measure advancing and
receding contact angles. The mean values across all the recorded data
are $\theta_a= 63.5\degr\pm3.7\degr$ and $\theta_r=8.2\degr\pm1.5\degr$.
Using these data, $\Delta(\cos\theta)_{a,r}=0.543\pm0.058$. Using the
$w_0$ data it is possible to calculate the contact angle of the initial
drop application in the low-Bond, spherical-cap limit. Using a fit to
the $(V,w_0)$ data for all drops below $175\ \mu$L yields a contact
angle of $50.8^{\circ}\pm4.8^{\circ}$, somewhat less than the measured
advancing contact angle. 

The height of the drop relative to the boundary layer thickness
determines the appropriate velocity scale for the $\We$ and $Re$. At
critical conditions, nearly all the $h_0$ values equal or exceed
$\delta_{99}$, the height in the boundary layer at which
$U(y)=0.99\,U_{\infty}$. Therefore, $U_{\infty}$ is used as the
reference velocity. Other constant physical parameters correspond to
conditions at $22^{\circ}$C: $\gamma = 0.0724$~N/m, $\rho_a 
= 1.20$~kg/m$^3$, $\rho_d = 998$~kg/m$^3$, and
$\mu_a = 18.2 \times 10^{-6}$~kg/m$\cdot$s.

\section{Depinning Limits}
\label{depinning}

\begin{figure}
\includegraphics[trim=48 48 160 496]{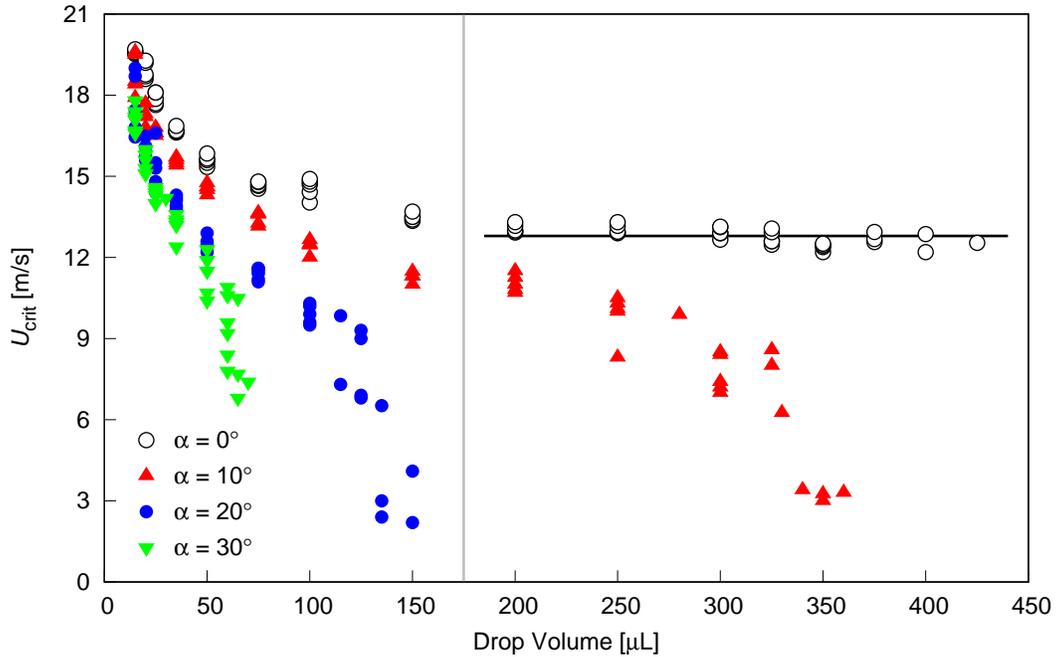}
\caption{Critical runback velocity, $U_{\mathrm{crit}}$ as a function
of drop volume, $V$. The vertical gray line is at $V=175\ \mu$L,
the boundary between the low- and high-Bond number regimes.}
\label{crit-U}
\end{figure}

To test the model represented by \eqn{critical-wind}, depinning
experiments were conducted using 220~drops ranging in volume from~15
to $425\ \mu$L and at surface inclination angles of $\alpha=0\degr$,
10\degr, 20\degr, and 30\degr. Data for critical depinning wind
velocity, $U_{\mathrm{crit}}$, is plotted as a function of drop
volume in Fig.~\ref{crit-U}. For small volumes, $U_{\mathrm{crit}}$
decreases rapidly as volume increases. As volume increases further,
drops on the $\alpha=0\degr$ surface reach a constant value,
$U_{\mathrm{crit}} \approx 12.8$~m/s indicated by the horizontal
black line. As inclination increases, the gravity force increases
and this reduces the wind velocity required for depinning.

\begin{figure}
\includegraphics[trim=48 48 310 564]{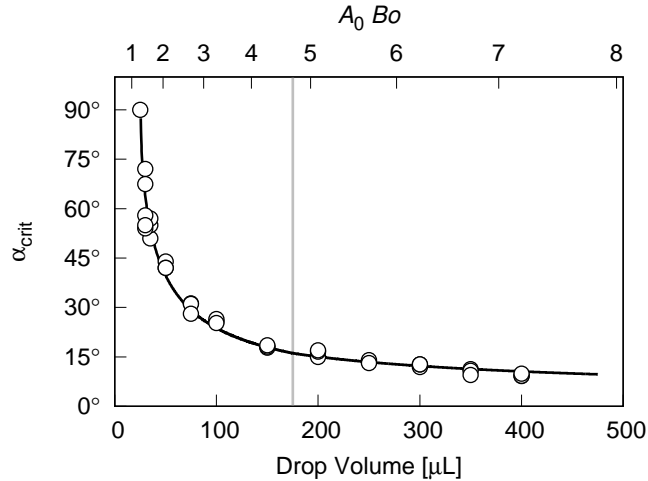}
\caption{Critical runback angle for gravity-forced drops, $U=0$.
The vertical gray line is at $V=175\ \mu$L, the boundary
between the low- and high-Bond number regimes.}
\label{crit-angles}
\end{figure}

To evaluate pure gravity-forced runback, a series of tests was
conducted in which the surface inclination was increased at a rate
of 1\degr\ per second until a critical runback angle was reached.
Results are shown in Fig.~\ref{crit-angles} with a best-fit curve
to Eqn.~(\ref{critical-tilt}). This fit yields $\Cgcrit = 1.323 \pm
0.013$ for this set of wettability conditions.  The fit is successful
across the entire range of \ABo\ in spite of $C_g^*$ varying above
$V=175\ \mu$L or $\ABo=4.76$. Because $C_g^*$ varies, \Cgcrit\ must
as well but the variations are sufficiently small to allow a
successful data fit.

For gravity-forced drops, the critical pinning force
is $\rho_d Vg\sin\alpha_{\mathrm{crit}}$. Using the
$\Cgcrit$ result from the inclined surface tests and the small-\Bo\
$C_g^*$ value yields $\Cgcrit^* = C_g^* \, \Cgcrit \approx 0.55$. This
value is expected to be close to $\Delta(\cos\theta)_{a,r}$ and the
match is outstanding for these experiments:
$\Delta(\cos\theta)_{a,r}=0.543\pm0.058$.

\begin{figure}
\includegraphics[trim=48 48 160 496]{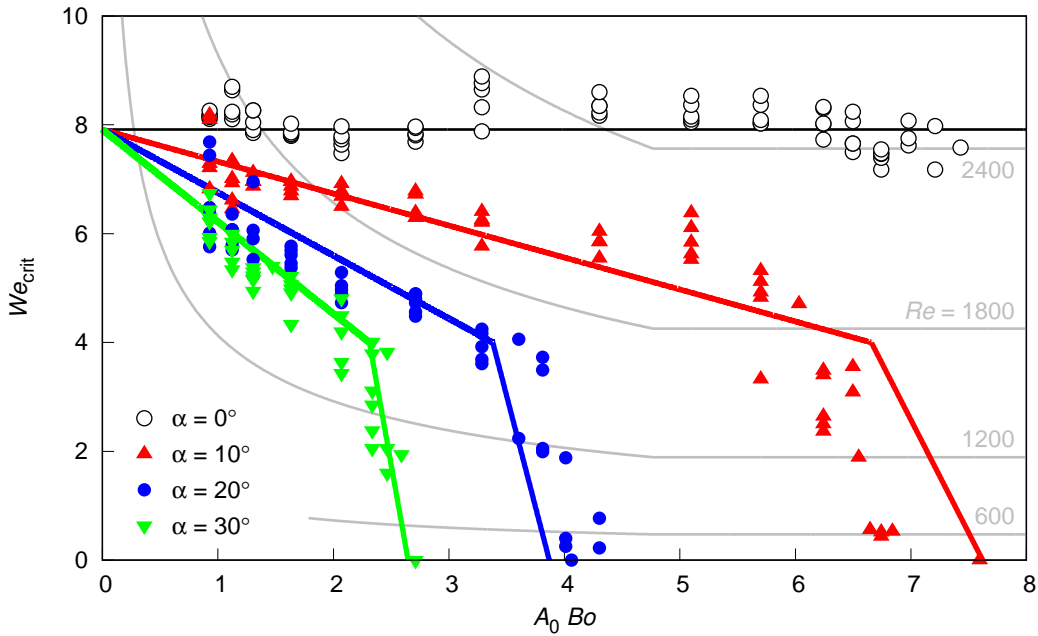}
\caption{Critical Weber number as a function of drop aspect ratio times Bond number.}
\label{crit-We}
\end{figure}

Returning to experiments with wind forcing, depinning threshold
data from Fig.~\ref{crit-U} is presented in nondimensional form in
Fig.~\ref{crit-We}. Data for the $\Wecrit=0$ limit is generated
using Eqn.~(\ref{critical-tilt}) with $\alpha_{\mathrm{crit}} =
10\degr$, 20\degr, and 30\degr. When the surface is horizontal,
$\alpha=0\degr$, drops run back at an essentially constant Weber
number, $\Wecrit = 7.9$.  This finding is consistent with the model
represented by Eqn.~(\ref{critical-wind}) and indicates that the
ratio $\Cgcrit\big/\Cwcrit$ is not a strong function of \ABo. This,
in turn, suggests that neither the drag coefficient nor the surface-tension
pinning coefficient are strong functions of the drop volume.

Large-volume drops subject to pure wind forcing demonstrate significant
interface oscillations, even at wind speeds below $U_{\mathrm{crit}}$.
For all but the smallest drops, $\alpha=0\degr$ critical conditions
occur above $Re=1800$ where significant vortex shedding is
expected~\cite{Acarlar-JFM-87}. The $\alpha=20\degr$ and 30\degr\ inclined-surface
tests show depinning at $Re<1800$ and drops in these tests do not undergo
significant interface-shape oscillations. While unsteady vortex shedding
likely occurs in the drop wakes at these Reynolds numbers, it seems
insufficiently strong to cause significant drop-shape unsteadiness.

For the three inclined-surface data sets, the critical Weber number
decreases linearly with increasing \ABo\ as predicted by
Eqn.~(\ref{critical-wind}) but does so in two distinct stages. Using the
combined data from all four surface inclination angles but restricting
to drops for which $\Wecrit>4$, Eqn.~(\ref{critical-wind}) is fit
successfully using $\Cgcrit = 2.338 \pm 0.056$ and $\Cwcrit = 0.2955 \pm
0.0083$. These coefficients apply to all four $\alpha$
values and to drops both in the small-\Bo\ range, $\ABo<4.76$
and in the large-\Bo\ range, $\ABo>4.76$. The ratio
$\Cgcrit\big/\Cwcrit$ yields $\Wecrit = 7.9$ for $\alpha=0\degr$.

Figure~\ref{crit-We} shows distinctly different slopes for drops with
$\Wecrit<4$ and this requires different values for \Cgcrit\ and
\Cwcrit. $\Wecrit<4$ will be referred to as the gravity-dominated forcing regime
while $\Wecrit>4$ will be referred to as the wind-dominated forcing regime.
$\Cgcrit$ in the gravity-dominated forcing regime is taken to be $1.323 \pm
0.013$, the value found in the critical-tilt experiments. Proceeding
from this, \Cwcrit\ is determined by requiring the curves be
continuous at $\Wecrit=4$. This yields $\Cwcrit = 0.042 \pm 0.017$.

The choice of $\Wecrit=4$ as a cutoff between the forcing regimes
provides the best match to the data. However, the quality of the fits
presented in Fig.~\ref{crit-We} are not especially sensitive to that
Weber number; values between 3.75 and 4.25 give essentially the same fit
quality. Notably, the different forcing regimes do not appear related to
drop volume. Both wind- and gravity-dominated depinning occurs at small
and large Bond numbers and the different critical coefficient values are
associate with different \Wecrit\ regimes, not different \ABo\ regimes.

The \Cwcrit\ coefficient in the gravity-dominated forcing regime has a larger
relative uncertainty than the other coefficients. This is reflected in
the worse data fit in the gravity-dominated regime, especially for
$\alpha=10\degr$. This may occur because the data in the
gravity-dominated regime is not used to determine the corresponding
\Cgcrit\ and \Cwcrit\ values. Again, the lines below $\Wecrit=4$ in
Fig.~\ref{crit-We} use the value of \Cgcrit\ from the critical tilt
tests plus the value of \Cwcrit\ that yields piecewise-continuous
extensions of the lines above $\Wecrit=4$. The relatively poor fit for
$\alpha=10\degr$ may also occur because the entire subset of the data for gravity-dominated forcing occurs at $\ABo>4.76$ where the initial drop shapes
are not spherical caps and \Cgcrit\ and \Cwcrit\ are expected to depend
at least weakly on \ABo.

To summarize these findings, the data and fit curves presented in
Fig.~\ref{crit-We} show that the simple mechanistic model of
mixed-mode depinning is successful but must be considered in distinct
wind- and gravity-dominated forcing regimes. The choice to represent the
data in two linear segments is arbitrary. A more sophisticated
model would include values of \Cgcrit\ and \Cwcrit\ that are
continuous functions of \Wecrit, \ABo, or other parameters. The
present model does not attempt to predict coefficient values as,
say, functions of $\theta_a$, $\theta_r$, Bond number, or Reynolds
number. However, once the coefficients are empirically determined, they
suggest the relative strengths of wind and pinning forces as
compared to the gravity force.

An unexpected result is that \Cgcrit\ is substantially larger in the
wind-dominated forcing regime as compared to the gravity-dominated regime:
$\Cgcrit = 2.338 \pm 0.056$ for wind-dominated forcing as compared to
$1.323 \pm 0.013$ for the gravity-only tests. The gravity-only value is
essentially equal to $\Delta(\cos\theta)_{a,r}$. This finding was
expected based on previous work dating to Macdougall and
Ockrent~\cite{Macdougall-PRSLA-42}. The fact that \Cgcrit\ is 75\%
larger for wind-dominated forcing means that surface tension is 75\%
more effective at resisting wind-dominated forcing than gravity-only
forcing. This suggests results from tilted plate tests may not be suitable for
predicting the maximum surface-tension pinning force in wind-forced situations. 

The different values of \Cgcrit\ and \Cwcrit\ above and below
$\Wecrit=4$ may arise because of different drop shapes in the wind- and
gravity-dominated forcing modalities. Different drop shapes also exist
in the small- and large-\Bo\ ranges below and above $\ABo=4.76$.
However, those changes are accommodated using constant critical
coefficients across the entire range of \ABo. Different contact-line
shapes and contact-angle distributions about the contact line would
affect \Cgcrit\ while different drop interface shapes would affect the
drag coefficient, \Cwcrit. To explore this possibility, data on drop
shapes under the different forcing modalities are presented in the next
section.

\section{Drop Shapes at Critical Depinning Conditions}
\label{shape}

To investigate why depinning conditions may differ between the
two forcing regimes, drop images were recorded using
side- and top-view cameras. Top-view images were analyzed using the
laser-speckle interface measurement technique developed by Schmucker
and coworkers~\cite{Schmucker-PhD-12,Schmucker-ExpFluids-12}. That
analysis consists of finding the contact line, measuring the
deformation of a laser speckle image caused by refraction of light
at the air/water interface, then numerically reconstructing the
interface shape using a simulated annealing optimization procedure.
Results are used to measure the contact-line shape and side-view
drop profile, both of which are measured with good accuracy.
Receding contact angles proved difficult to measure with good accuracy
using this approach. Receding and advancing contact angles
reported in Sec.~\ref{setup} were measured using conventional
side-view images and are accurate to within the quoted uncertainty.

\begin{figure}
\includegraphics{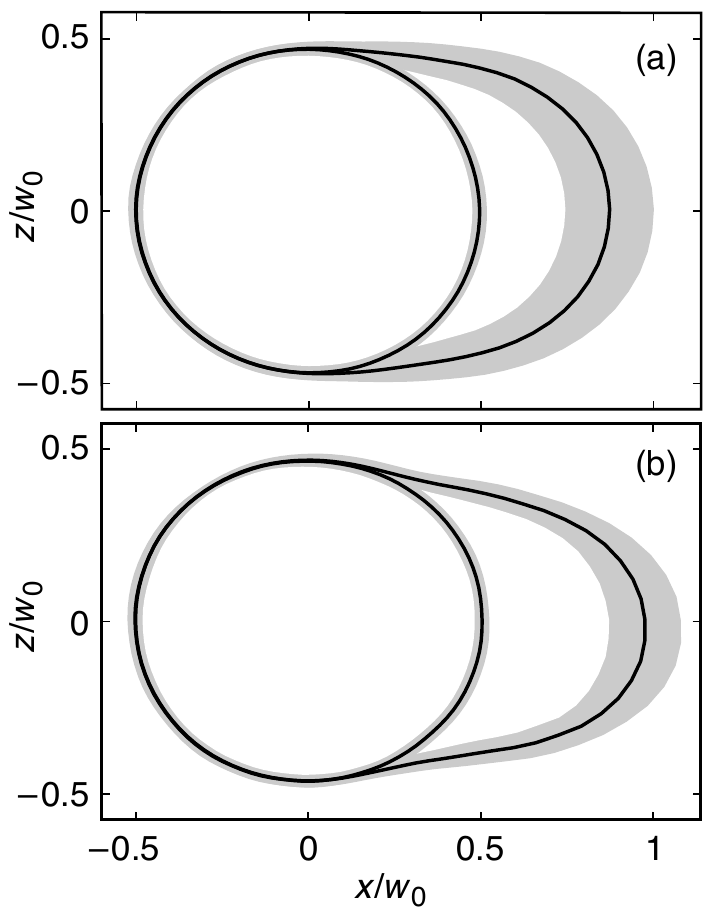}
\caption{Mean and standard deviation of drop contact lines for initial
drop placements and final subcritical drop measurements. Contact lines
are shown for (a) wind-forced drops and (b) gravity-only drops. Solid
curves are mean contact line locations. Shaded areas indicate one
standard deviation.}
\label{CL-Composite}
\end{figure}

Mean initial and final contact line shapes for drops
that depin due to mainly wind forcing are shown in
Fig.~\ref{CL-Composite}(a) while the equivalent mean shapes for the
gravity-only drops are shown in Fig.~\ref{CL-Composite}(b). Each
drop's geometry is scaled by its initial width in the streamwise $x$ direction, $w_0$.
Although the differences between gravity-only forcing and wind-dominated
forcing are small, drops subjected to gravity-only forcing elongate
somewhat more than the wind-forced drops. The most-downstream
point of the gravity-only drops is $0.98w_0$ from the initial drop
center as compared to $0.87w_0$ for the wind-forced drops.
Additionally, the radius of curvature at the advancing part of the
contact line is smaller for gravity-forced drops, $0.31w_0$, than
for wind-forced drops, $0.39w_0$. The drops' maximum transverse
width was not observed to decrease for either forcing type.

\begin{figure}
\includegraphics{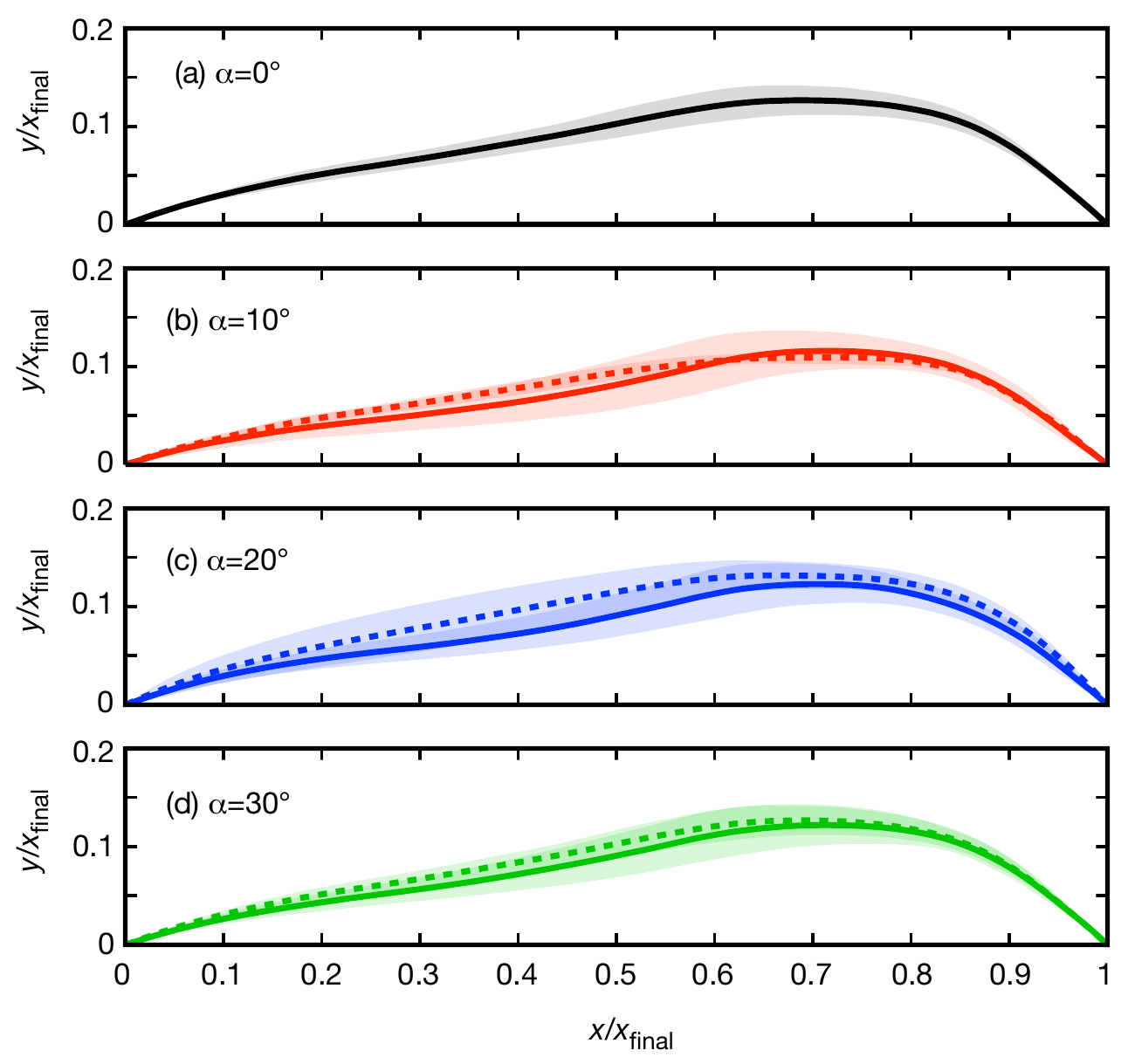}
\caption{Final subcritical drop profiles for each surface inclination
angle. Mean profiles for drops that depin in the wind-dominated forcing regime
are shown with solid lines; drops that depin due to gravity-dominated forcing are
shown with dashed lines. Shaded areas indicate one standard deviation.}
\label{Profile-Composite}
\end{figure}

Sideview profiles of the final subcritical drop configurations are
shown in Fig.~\ref{Profile-Composite}. The plots are scaled by each
drop's final streamwise length. Dashed lines are drops that depin
in the gravity-dominated forcing regime; solid-lines represent wind-dominated forcing.  
Concave interface curvature is present on the windward portion of the drop interface
at high wind speeds.  This shows the wind pressure imposed on that
portion of the drop exceeds the capillary pressure that would be
present if the drop was able to maintain its spherical cap shape.
Again, the receding contact angles are not captured accurately using
the top-view laser speckle technique; the receding contact angles
are larger than what is observed in the side-view profiles.

One difference between the forcing regimes that
is not highlighted by Fig.~\ref{Profile-Composite} is the difference
in final drop length and height. Gravity-forced drops are approximately
13\% longer than same-volume wind-forced drops just before
depinning. Because the side-view profiles are nearly unchanged for
the two regimes, this means the maximum drop height is larger by
approximately the same amount. The larger height might be expected
to result in a larger value of \Cwcrit\ (the drag coefficient) for
primarily gravity-forced drops as compared to primarily wind-forced drops. However,
this is not the case. Wind-forced drops have a substantially
larger value of \Cwcrit.  This
could be a consequence of interface shape differences or the increased
interface unsteadiness at higher Reynolds numbers.

Ultimately, the reasons for the different \Cgcrit\ and \Cwcrit\
values are not clear. Drops do exhibit different contact line shapes and
profiles just prior to depinning in the two forcing regimes.
However, the differences do not suggest an obvious
connection to the different critical coefficient values. To provide
more information on this issue, higher-accuracy measurements of the
contact angle distribution about the contact line are needed.

\section{Conclusions}
\label{conclusions}

The objective of this work is to develop and test a simple mechanistic
model of gravity- and high-$Re$ wind-forced drop depinning using
straightforward formulations of surface-tension, gravity,
and aerodynamic forces. To test the model, depinning experiments
were performed using water drops ranging from~15 to 450~$\mu$L on
a roughened aluminum surface inclined at 0\degr, 10\degr, 20\degr,
and 30\degr\ in combination with wind forcing. Tests of critical
tilt angle without wind forcing were also performed. Critical tilt
experiments yield the critical pinning-force coefficient \Cgcrit\ for
gravity-only forcing. Tests with wind-only forcing on the non-inclined
surface resulted in a constant critical Weber number. Combined
wind and gravity forcing experiments give critical pinning and
drag coefficients under mixed forcing conditions. Overall, the model
is judged to be successful.

When the surface is inclined, the model predicts the critical Weber
number decreases linearly with increasing \ABo, the product of the
initial drop aspect ratio and Bond number. The unknown model
coefficients \Cgcrit\ and \Cwcrit\ do not depend strongly on either
Bond number or surface inclination. However, they do take markedly
different values above and below $\Wecrit = 4.0$ which is identified
as a boundary between wind- and gravity-dominated forcing regimes for the
present wettability conditions. The existence of different regimes
has not been previously identified and has important implications
for predicting drop depinning due to wind forcing using data obtained
from critical tilt experiments.

Contact line shape and side-yield profile shapes were measured in
an attempt to explain the different values of \Cgcrit\ and \Cwcrit\
in the different forcing regimes. In both forcing modalities, the
contact lines of the final drop shapes appear as semicircular arcs
connected by nearly straight-line segments. Compared to drops that depin
in the wind-dominated forcing regime, gravity-forced drops extend
somewhat more in the streamwise direction before depinning and their
contact lines have a smaller radius of curvature on the advancing side.

Overall, the difference in the model coefficients in the wind- and
gravity-regimes is not clearly explained by the shape measurements
presented here. Continuing work aims to improve receding contact
angle measurements and this may provide additional information that
could improve understanding. Separately, the role of Reynolds
number and drop interface unsteadiness remains under active
investigation.

\newpage

\section*{Acknowledgements}

The authors wish to thank Sungyon Lee, Alireza Hooshanginejad, and
Roger Simon for multiple fruitful conversations regarding this work.
The authors also with to thank the U.S. National Science Foundation
for support from grants CBET-0828469 and CBET-1839103.

\bibliography{White-PRF-2x-Manuscript}

\end{document}